\documentclass[aps,reprint,amsmath,amssymb,superscriptaddress,pra]{revtex4-1}

\usepackage{dsfont}
\usepackage{graphicx}
\usepackage{subfigure}
\usepackage{amsmath}
\usepackage{braket}
\usepackage{float}
\usepackage[colorlinks]{hyperref}
\hypersetup{
	colorlinks = true,
	urlcolor = {blue},
	citecolor = {blue},
	linkcolor= {blue}
}
\usepackage{slashbox}

\DeclareMathOperator{\Tr}{Tr}

\renewcommand{\eqref}[1]{(\ref{#1})}
\newcommand{\figref}[1]{Fig.~\ref{#1}}
\newcommand{\appref}[1]{App.~\ref{#1}}
\newcommand{\secref}[1]{Sec.~\ref{#1}}

\newcommand{\red}[1]{{\color{red}#1}}

\begin{document}

\title{Improving semi-device-independent randomness certification by entropy accumulation}

\author{Carles Roch i Carceller}\email{carles.roch\_i\_carceller@teorfys.lu.se}\address{Center for Macroscopic Quantum States (bigQ), Department of Physics,
Technical University of Denmark, 2800 Kongens Lyngby, Denmark}\address{Department of Physics and NanoLund, Lund University, Box 118, 22100 Lund, Sweden}

\author{Lucas Nunes Faria}\address{Center for Macroscopic Quantum States (bigQ), Department of Physics,
Technical University of Denmark, 2800 Kongens Lyngby, Denmark}

\author{Zheng-Hao~Liu}\address{Center for Macroscopic Quantum States (bigQ), Department of Physics,
Technical University of Denmark, 2800 Kongens Lyngby, Denmark}

\author{Nicolò Sguerso}\address{Center for Macroscopic Quantum States (bigQ), Department of Physics,
Technical University of Denmark, 2800 Kongens Lyngby, Denmark}

\author{Ulrik~Lund~Andersen}\address{Center for Macroscopic Quantum States (bigQ), Department of Physics,
Technical University of Denmark, 2800 Kongens Lyngby, Denmark}

\author{Jonas Schou Neergaard-Nielsen}\address{Center for Macroscopic Quantum States (bigQ), Department of Physics,
Technical University of Denmark, 2800 Kongens Lyngby, Denmark}

\author{Jonatan Bohr Brask}\email{jonatan.brask@fysik.dtu.dk}\address{Center for Macroscopic Quantum States (bigQ), Department of Physics,
Technical University of Denmark, 2800 Kongens Lyngby, Denmark}

\begin{abstract}
Certified randomness guaranteed to be unpredictable by adversaries is central to information security. The fundamental randomness inherent in quantum physics makes certification possible from devices that are only weakly characterised i.e.\ requiring little trust in their implementation. It was recently shown that the amount of certifiable randomness can be improved using the so-called Entropy Accumulation Theorem generalised to prepare-and-measure settings. Furthermore, this approach allows a finite-size  analysis which avoids assuming that all rounds are independent and identically distributed. Here, we demonstrate this improvement in semi-device-independent randomness certification from untrusted measurements.
\end{abstract}

\maketitle

\section{Introduction}

Randomness is a fundamental resource \cite{hayes2001} for a variety of applications in the modern world including information security \cite{shannon1949,Niels2011}, simulations of physical systems \cite{MonteCarlo1949,Metropolis1953,Montanaro2015,Ghersi2017,Miyamoto2020}, games, and gambling. Cryptographic protocols in particular require unpredictability relative to potential adversaries of the random numbers used to generate keys.

For pseudo-random number generators \cite{james1990} and randomness generated from classical physical processes, certification of unpredictability requires assumptions about the information
and computational power available to the adversaries. The inherent randomness of quantum measurements, on the other hand, enables randomness certification directly from fundamental physical laws and measurable properties of the devices used \cite{ma2016,herrero2017,bera2017,grangier2018}. From a characterisation of states and measurements on a given system, the entropy (quantifying unpredictability) relative to any adversary constrained by quantum mechanics can be bounded. In fact, by exploiting quantum nonlocality \cite{brunner2014}, the need for a thorough characterisation can even be eliminated, allowing randomness certification in a black-box setting, provided the devices violate a Bell inequality \cite{colbeckPhD2009,pironio2010,acin2016}. This corresponds to a very strong level of security, known as device independence (DI), since minimal trust in the devices is required \cite{Christensen2013,Liu2018,Bierhorst2018,Shalm2021,Liu2021}. DI schemes, however, are also more technologically challenging to implement than device-dependent ones with well characterised devices. Hence, it is desirable to identify good trade-offs between ease of implementation and how little trust is required to certify randomness, i.e.\ to explore semi-DI  quantum random number generators (QRNGs). A number of different prepare-and-measure schemes with fully uncharacterised sources \cite{vallone2014,Cao2016,Xu2016,marangon2017,Avesani2018,Michel2019,Drahi2020,lin2020,lin2022,xing2022} or measurements \cite{passaro2015,Chaturvedi2015,cao2015,Nie2016,bischof2017,pivoluska2021,Wang2023,Argillander2023,lin2023} have been explored. Other approaches present semi-DI protocols restricted by dimension \cite{li2011,lunghi2015,mironowicz2021}, energy \cite{himbeeck2017,rusca2019,himbeeck2019,tebyanian2021}, contextuality \cite{carceller2022} or overlaps between the state preparations \cite{brask2017}. The semi-DI protocol presented in work fits into the latest, i.e.\ the considered state preparations are partially characterised with a bound on their overlaps, leaving the measurement device as a black-box.

Randomness in a variable $b$ with respect to an adversary Eve is typically certified by bounding the capacity of this adversary of guessing $b$. Here, we look at a semi-DI prepare-and-measure scenario assuming only pure state preparations with bounded overlaps and untrusted measurements with $b$ as measurement outcomes. Considering that Eve has bounded knowledge and control over the setup used to generate $b$, one can find certifiable bounds on Eve's guessing probability $p_g$ which can be used to compute the min-entropy $H_{min}=-\log_{2}(p_g)$ \cite{konig2009} of the measurement record. While the single-round min-entropy has served as a benchmark quantifier of randomness, it has been shown that one can bound the min-entropy of a sequence of outcomes in terms of the single-round Shannon entropy instead if the experiment runs over a large number of rounds \cite{tomamichel2009}. This provides more randomness but the statement is valid only under the unrealistic assumption that all rounds are independent and identically distributed (i.i.d.). However, it has recently been shown that one can circumvent the i.i.d.~limitation \cite{arnon2018,metger2022,metger2023,zhou2023}, in particular using the so-called Entropy Accumulation Theorem (EAT) generalised to prepare-and-measure scenarios \cite{metger2022,metger2023}. Although its promising breakthrough in DI quantum cryptography, the EAT has found limited practical applicability in semi-DI QRNG frameworks. The main reason lies in the difficulty of finding the suitable mathematical tools known as \textit{min-tradeoff} functions required apply the EAT \cite{brown2021}. These are arbitrary affine functions on the observable frequencies which lower-bound the von Neumann entropy. The difficulty of finding min-tradeoff functions boils down to computing affine lower-bounds on a non-linear quantity such as the von Neumann entropy. Many works address this issue by computing the bounds on the min-entropy, which is far more accessible through semidefinite program techniques. Here, we address this issue by computing affine lower-bounds on the Shannon entropy based on semidefinite programming duality.

In this work, we demonstrate a simple semi-DI QRNG to showcase the power of the EAT in prepare-and-measure scenarios. Based on semidefinite program duality, we present a method to produce min-tradeoff functions and together with the generalised version of the EAT applicable to prepare-and-measure scenarios, deal with finite size effects without assuming i.i.d. rounds. Here, we aim to certify randomness versus classically correlated eavesdroppers. Indeed, fundamental bounds on certifiable randomness require to consider eavesdroppers with access to the most powerful tools available in nature, i.e.~quantum correlations. In reality however, harnessing non-local correlations from long-lived entanglement is extremely difficult with current technologies. 
Moreover, even if entanglement can be successfully generated, it needs to be stored in order to gather accumulated information after each experimental round. Thus, if the absence of quantum memories within the devices can be verified, quantum side information can be straightly ruled out \cite{knill2020}. Here we assume we find ourselves in that case: we only allow classical side information in the sense that our randomness bounds are only valid for classically correlated eavesdropping entities. In the end, assuming the nature of the side information in a quantum randomness certification protocol is a matter of security setting. In quantum cryptography, for instance, with spatially separated communicating parties, a malicious third party has a wider margin to become entangled compared to a prepare-and-measure QRNG scenario, where all the action can be closed up in a microchip.

Our protocol is able to certify more than one bit of randomness per round from measurements on a single quantum state. We implement the scheme in a setup using time-bin encoded states and single-photon detection. The protocol employs a prepare-and-measure setup with three states and a single three-outcome measurement. The scheme is semi-DI in the sense that the only assumption on the source is a bound on the pair-wise overlap of the prepared states, and the measurement is completely uncharacterised. Randomness is extracted from one of the inputs, and the amount of certified randomness achievable with the generalized EAT is compared with the traditional min-entropy approach. The other two inputs enable self-testing the device, by exploiting that the setup can also be interpreted in terms of a state discrimination task. 

In quantum state discrimination \cite{bergou2007,barnett2009,bae2015}, a measurement device aims to determine which state out of a known set was prepared, subject to certain constraints. In unambiguous state discrimination (USD) \cite{ivanovic1987, dieks1988, peres1988}, the error rate is nullified, at the cost of adopting an additional measurement event which gives no information on which state was prepared (formally called an inconclusive event). The minimum attainable rate of inconclusive events in USD is given by a rank-1 (hence extremal) and unique measurement, which benefits the amount of certifiable randomness on the measurement outcome. While a setup with two preparations ($x=0,1$) is sufficient for QRNG, we add a third state preparation ($x=2$) such that all outcomes are equally distributed when Alice selects $x=2$. The key idea is then that, to be able to reproduce the observable statistics for inputs $x=0,1$, Eve must perform the USD measurement. Therefore, when Alice selects $x=2$, Eve’s chances of guessing the outcome become minimal. Here, we implement a protocol which uses USD during the self-test rounds, allowing us to make use of the additional inconclusive event in two-state discrimination to reach randomness values greater than one bit per round. Then, we choose a third state that yields equiprobable outcomes when involved in the state discrimination scenario. 

The remainder of the paper is organised as follows. In \secref{sec.results}, we introduce the state discrimination scenario and specify the measurement strategy. Later, we explain how we evaluate the randomness, continue with the main semi-device independent assumptions we consider and explain how we deal with finite size effects assuming independent and identically distributed (i.i.d.)~and non-i.i.d.~rounds. We end the section by experimentally implementing the protocol in an optical platform using coherent states of light. In section \secref{sec:results} we present the results we observe both in the experiment and in simulations. We end the paper in \secref{sec.discussion} concluding our work and outlining the main result. 

\begin{figure}
    \centering
    \includegraphics[width=0.49\textwidth]{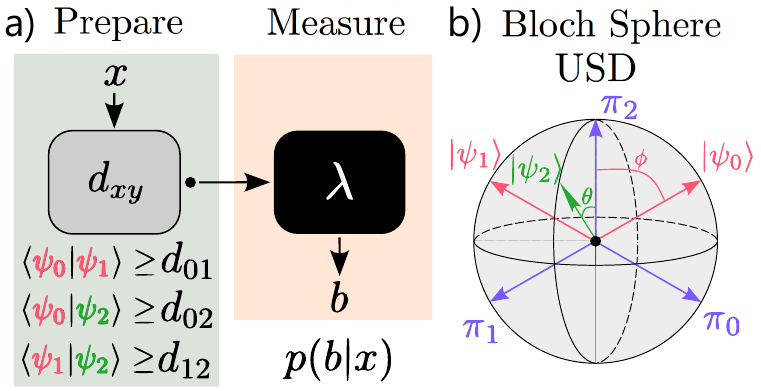}
    \caption{\textbf{a)} Sketch of the semi-device independent prepare-and-measure scenario with three preparations and one untrusted measurement (black box). \textbf{b)} The USD protocol is illustrated on the Bloch sphere, with Bloch vectors corresponding to the prepared states $\ket{\psi_x}$  (red and green) and POVM elements $\pi_b$ (blue).}
    \label{fig:1}
\end{figure}

\section{Methods}
\label{sec.results}

\subsection{Prepare-and-measure}

In order to illustrate the main idea of the protocol, consider the scenario in \figref{fig:1}a, where one party (namely Alice) owns a device that receives inputs $x\in\left\{0,1,2\right\}$ with prior probabilities $p_x$ and prepares the following pure quantum states,
\begin{align}
\label{eq:states10}
    &\ket{\psi_0} = \cos\frac{\phi}{2}\ket{0} + \sin\frac{\phi}{2}\ket{1} \\
    &\ket{\psi_1} = \cos\frac{\phi}{2}\ket{0} - \sin\frac{\phi}{2}\ket{1} \ , \nonumber
\end{align}
and a third qubit state $\ket{\psi_2}$ which shall remain unspecified for now. These states are sent to a second party (namely Bob) who owns a device which will perform a measurement described by the POVM $\left\{\pi_{b}\right\}$ for $b\in\left\{0,1,2\right\}$. Over many rounds of the experiment, one can estimate the conditional probability distribution $p(b|x)=\Tr\left[\rho_x \pi_b\right]$, for $\rho_{x}=\ket{\psi_{x}}\bra{\psi_{x}}$.

Our first task is to find Bob's optimal measurement, according to a particular state discrimination protocol, to identify whether Alice prepared $x=0$ or $x=1$, ignoring the third possible input $x=2$. This will yield correlations reproducible only with extremal and unique POVMs. The idea is to properly design a state $\ket{\psi_2}$ such that, with that optimal measurement, all outcomes are equiprobable whenever that state is prepared (i.e. $p(b|x=2)$ are equal $\forall b$). We take optimality to be set by USD, i.e.~a measurement that minimises the rate of inconclusive events (here labeled by $b=2$) given zero error rate. Correlations drawn from optimal USD measurements are extremal and unique, which means that they cannot be decomposed by any convex combination of other POVMs and no other measurement can reproduce the same probabilities, up to unitary rotations.

\subsection{Unambiguous state discrimination}

The task in USD is to identify which state was prepared without making any errors, i.e.~$p_{\text{error}}:=p_0 p(1|0)+p_1 p(0|1)=0$ \cite{ivanovic1987,peres1988,dieks1988}. That can be done if one pays the price of having some rounds in which the measurement result turns inconclusive. In the present case, USD targets preparations $x=0$ and $x=1$, and the inconclusive events will be labeled with $b=2$. The goal of USD is to minimize the rate of inconclusive events $p_{\text{inc}}:=p_0 p(2|0) + p_1 p(2|1)$. In two state discrimination, the minimum $p_{\text{inc}}$ is proportional to the overlap of the prepared states, which in our scenario is $|\braket{\psi_0|\psi_1}|=\cos\phi$, according to \eqref{eq:states10}. In this case, the minimum $p_{\text{inc}}$ is lower bounded by \cite{carlesPhD2023}
\begin{align}
    p_{\text{inc}} \geq 2\sqrt{p_0 p_1} \cos\phi \ .
\end{align}
The POVM that represents an optimal USD measurement must be given by rank-1 POVM elements, proportional to the projectors onto the orthogonal states of $\ket{\psi_0}$ and $\ket{\psi_1}$. Concretely, for equiprobable preparations $p_0=p_1=1/2$,
\begin{align}
    \pi_0 &= \frac{1}{1+\cos\phi}\ket{\psi_1^{\perp}}\bra{\psi_1^{\perp}} \nonumber \\
    \pi_1 &= \frac{1}{1+\cos\phi}\ket{\psi_0^{\perp}}\bra{\psi_0^{\perp}} \label{eq:POVM_USD} \\
    \pi_2 &= \mathds{1} - \pi_0 - \pi_1 \nonumber \ ,
\end{align}
where $\braket{\psi_x|\psi_x^{\perp}}=0$. Consider now a third preparation $x=2$. We aim to find a state $\ket{\psi_2}$ which triggers all three outcomes $b=0,1,2$ of the measurement in \eqref{eq:POVM_USD} with the same probability. That means, it must satisfy $\bra{\psi_2}\pi_{b}\ket{\psi_2}=1/3$ $\forall b$. As illustrated on the Bloch sphere in \figref{fig:1}b, a good candidate is the state
\begin{align}
\label{eq:third_state}
    \ket{\psi_2} = \cos\frac{\theta}{2}\ket{0} + i\sin\frac{\theta}{2}\ket{1}  \ .
\end{align}
The three outcomes will be equiprobable when 
\begin{align}
\label{eq:varphi_USD}
    \cos\theta = \frac{1-2\cos\phi}{3\cos\phi} \ .
\end{align}
Since $-1\leq\cos\theta\leq 1$, this condition can only be satisfied for $\cos\phi \geq 1/5$, meaning that equiprobable outcomes in this setting are only achievable in that range. One can then use these equiprobable outcomes, which are only reproducible through a unique and extremal POVM in a qubit space, to improve the randomness certification. Note, though, that this is only true for qubit states. In semi-DI randomness certification, one wants to keep the number of assumptions at minimum. We will later show how we can get rid of the assumption of a fixed dimension.

\subsection{Randomness certification} \label{sec:rng}

We proceed to explain how we certify randomness in the measurement outcomes, relative to a classically correlated adversary Eve. Eve has knowledge of the preparation ($x$) and measurement device, and can share classical correlations with the measurement according to the distribution $q(\lambda)$. We begin explaining how we certify single-round randomness using the min-entropy. Then, we bound the Shannon entropy in order to leverage the power of the EAT to accommodate the finite-size effects from experimental data. \\

\textbf{Min-entropy:} We certify randomness only from rounds where state $\ket{\psi_{2}}$ is prepared. The rounds where states $\ket{\psi_0}$ and $\ket{\psi_1}$ are prepared can be thought of as self-test rounds. During the rounds where $\ket{\psi_{2}}$ is prepared, Eve's guessing probability averaged through each round can be written as
\begin{align}
\label{eq:pg}
    p_{g} = \sum_{\lambda}q(\lambda)\underset{b}{\max}\left\{p(b|x=2,\lambda)\right\} \ ,
\end{align}
for $p(b|x,\lambda)=\Tr\left[\rho_x \pi_{b}^{\lambda}\right]$ ($\rho_x=\ket{\psi_x}\bra{\psi_x}$) being the probability that the measurement outcome is $b$ given state preparation $x$ and the measurement strategy $\lambda$.

We aim to find an upper bound on $p_{g}$ in \eqref{eq:pg} over all strategies of Eve, i.e.~all distributions $q(\lambda)$ and measurements $\pi_{b}^{\lambda}$, subject to reproducing the observed statistics on average $p(b|x)=\sum_{\lambda}q(\lambda)\Tr\left[\rho_{x}\pi_{b}^{\lambda}\right]$. This optimisation problem can be rendered as a semidefinite program \cite{pauldaniel2023}. As we detail in \cite{supp_mat}, we can write Eve's guessing probability as
\begin{align}
\label{eq:pg_sdp}
    p_{g} = \sum_{\lambda}\Tr\left[\rho_{2}M_{\lambda}^{\lambda}\right] \ .
\end{align}
An upper bound $p_{g}^{*}\geq p_{g}$ can be found by maximizing \eqref{eq:pg_sdp} over all positive semidefinite $D\times D$ matrices $M_{b}^{\lambda}$ that fulfil the constraints
\begin{align}
    &\sum_b M_b^\lambda = \frac{1}{D}\Tr\left[\sum_bM_b^\lambda\right]\mathds{1} \ \forall \lambda \ , \label{eq:cts_primal1} \\
    &p(b|x)=\sum_\lambda\Tr\left[M_b^{\lambda}\rho_x\right] \ \forall b,x \ . \label{eq:cts_primal2}
\end{align}
where $D$ is the eavesdropper's dimension. The randomness of the measurement outcomes is quantified through the min-entropy $H_{\rm min}=-\log_{2}\left(p_{g}\right)$, which gives the number of (almost) uniformly random bits which can be extracted per round of the protocol \cite{konig2009}. \\

\textbf{Shannon entropy:} Due to the non-linear nature of the Shannon entropy, it is very difficult to find linear schemes that provide tight numerical bounds in single-party prepare-and-measure scenarios. For example, in Ref.~\cite{himbeeck2019} the authors design a hierarchy of semidefinite programs to bound the Shannon entropy in scenarios with energy-restricted correlations. Here, our correlations are bounded by a constraint on the overlaps and thus, we find the following method more suitable. As shown in Ref.~\cite{brown2023}, applying the Gauss-Radau quadrature to an integral representation of the logarithm yields a variational upper bound for the quantum relative entropy $D\left(\rho||\sigma\right):=\Tr\left[\rho\left(\log\rho-\log\sigma\right)\right]$. This can be related to the von Neumann entropy, and consequently to the Shannon entropy for classically correlated eavesdroppers. In \cite{supp_mat} we show how to use this technique to derive the bound on the conditional Shannon entropy $H(B|E,x=2)\geq H^\ast$, for
\begin{align}
    H^\ast = c_m + \sum_{i=0}^{m-1} \ \max_i \left\{\sum_{b,x}-\nu_{bx}p(b|x) - \Tr\left[R\right]\right\} \label{eq:H_bound}
\end{align}
where we introduced $c_m:=\sum_{i=1}^{m-1}\tau_i$ with $\tau_i:=\omega_i/(t_i\log(2))$ for $\left\{t_i,\omega_i\right\}$ being the nodes and weights respectively, corresponding to the Gauss-Radau quadrature. The maximisation runs for each iteration of $i$ over the set of scalars $\nu_{bx}$ and $D\times D$ matrices $R$ that satisfy 
\begin{align}
& \sum_{a} E_{b}^{a} = \sum_{x}\rho_x \nu_{bx} 
     + R    \\
& F_{b}^{a} + F_{b}^{a\dagger} = 2\tau_i\rho_{2}\delta_{a,b} + Q^{a} - \frac{1}{d}\Tr\left[Q^{a}\right]\mathds{1}  \\
& L_{b}^{a}\!=\!\tau_i\rho_{x^\ast}\!\left[\left(1\!-\!t_i\right)\delta_{a,b}\!+\!t_i\right]\! +\! P^{a}\! -\! \frac{1}{d}\Tr\left[P^{a}\right]\mathds{1}   \ ,
\end{align}
for $Q^{a}$ and $P^{a}$ being $D\times D$ matrices, and $E_{b}^{a}$, $F_{b}^{a}$ and $L_{b}^{a}$ being blocks of $D\times D$ matrices that built the following positive semidefinite matrix
\begin{align}
 \begin{pmatrix}
        E_{b}^{a} & F_{b}^{a} \\
        F_{b}^{a\dagger} & L_{b}^{a}
    \end{pmatrix} \succeq 0 \ .
\end{align}
Any feasible point that fulfills all constraints specified above represents a valid lower bound on the Shannon entropy.

\subsection{Semi-device-independence}

We consider frameworks with the lowest device characterisation as possible in a semi-device-independent manner. Concretely, we consider that the measurement device is completely uncharacterised, and assume pure state preparations with known overlaps. For two pure state preparations, this allows using two fixed qubit states in the semidefinite program without loss of generality, because unitary rotations of the state pair will not affect the optimum, and Eve will not gain any extra information by extending the measurements beyond the two-dimensional span of the pair \cite{rusca2019}. In our case we consider three pure states, which means that Eve's maximal useful dimensionality will not exceed a qutrit space. Hence, we need to assume that the three prepared states span a three-dimensional space. That is, the third state is
\begin{align}
\label{eq:third_tilde}
    \ket{\tilde{\psi}_2} = \sqrt{a}\ket{\psi_2} + \sqrt{1-a}\ket{2} \ ,
\end{align}
for $\ket{\psi_2}$ in \eqref{eq:third_state} having support on the bi-dimensional space spanned by $\ket{\psi_0}$ and $\ket{\psi_1}$ in \eqref{eq:states10}, and $\ket{2}$ has only support on an additional orthogonal dimension, such that $\braket{\psi_x|2}=0$ $\forall x$. One can see that, if both overlaps $|\braket{\tilde{\psi}_2|\psi_0}|^{2}$ and $|\braket{\tilde{\psi}_2|\psi_1}|^{2}$ are simultaneously fixed to be
\begin{align}
    |\braket{\tilde{\psi}_2|\psi_0}|^{2}=|\braket{\tilde{\psi}_2|\psi_1}|^{2} = \frac{1}{2}\left(1+\cos\phi\cos\theta\right)  \ ,
\end{align}
then the normalisation in \eqref{eq:third_tilde} imposes $a = 1$, and all three preparations must have support on a qubit space. However, strict equalities can be hard to satisfy in the lab. In \cite{supp_mat} we show how one can relax this assumption to lower bounds on the overlaps $|\braket{\psi_{x}|\psi_{y}}|\geq d_{xy}$. This, in turn, bounds the access of the eavesdropper to an additional third dimension,  as $a$ fulfills
\begin{align}
\label{eq:a_bounding}
    a\geq\frac{d_{02}^{2}+d_{12}^{2}-2d_{01}d_{02}d_{12}}{1-d^{2}_{01}} \ .
\end{align}
Also, since $a\leq 1$, one also finds the following relation
\begin{align}
\label{eq:delta_relation}
    1\geq d_{02}^{2}+d_{12}^{2}+d_{01}^{2}-2d_{01}d_{02}d_{12} \ ,
\end{align}
which must hold true for any dimension. Equation \eqref{eq:a_bounding} defines the surface of an inflated tetrahedron with curved faces (see \cite{supp_mat}). The amplitude $a$ decreases towards the center of the tetrahedron. Thus, in other words, bounding $|\braket{\psi_{x}|\psi_{y}}|\geq d_{xy}$ implies that we forbid Eve to access the interior of the tetrahedron.

This semi-DI characterisation allows us to compute the certifiable randomness beyond the two-qubit case and avoid direct dimensional constraints. In our implementation, we employ coherent states of light as pure state preparations. We control the pairwise overlaps of the preparation and use \eqref{eq:a_bounding} to bound their support outside a two-dimensional subspace.

\begin{figure*}
    \centering
    \includegraphics[width=\textwidth]{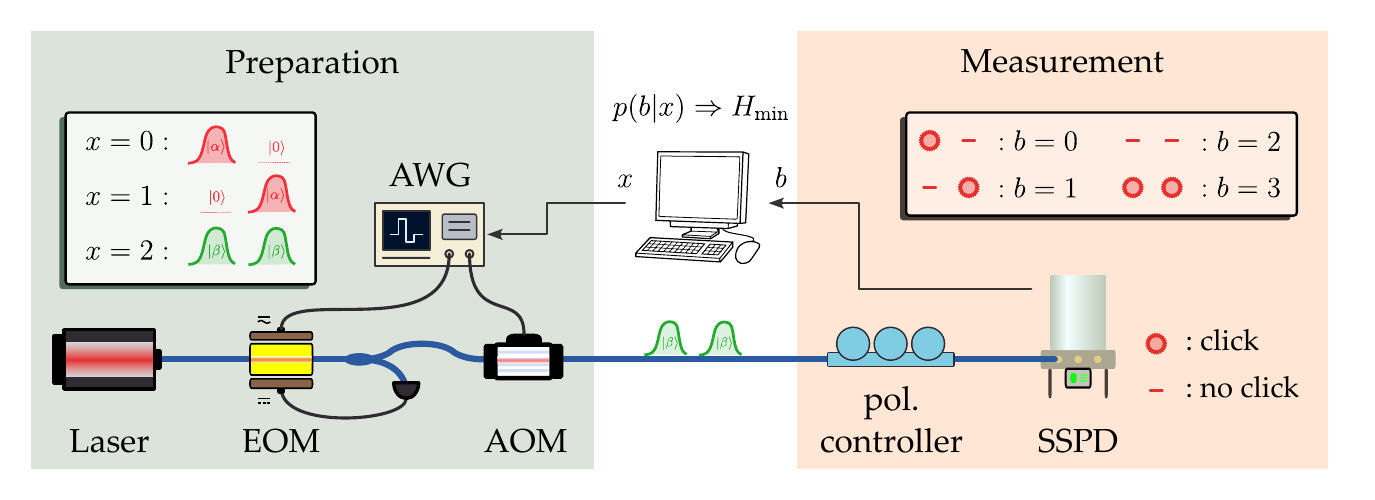}
    \caption{Experimental implementation. The time-bin-encoded input states are generated by an electro-optic modulator (EOM) and a pair of acousto-optic modulators (AOMs, only one shown in the figure). The feedback control at the EOM stabilizes the amplitude of the pulsed weak coherent states. After optimizing the polarization (pol.\ controller), the pulses are sent to the superconducting single-photon detector (SSPD). From the statistics of preparations and detector counts, randomness is extracted and certified.}
    \label{fig:implementation}
\end{figure*}

\subsection{Finite-size effects -- i.i.d.\ vs.\ entropy accumulation }

In actual experiments, probabilities are estimated from frequencies. Due to the finite number of data points, the observed frequency of events $\text{freq.}(b|x)=n_{bx}/n_{x}$, for $n_{bx}$ denoting the total number of events $b$ given a state preparation $x$ does not exactly represent the true conditional probability $p(b|x)$. To deal with finite-size effects, we take two different approaches. 

First, we assume that all rounds from the experiment are considered to be independent and identically distributed (i.i.d.). Under this assumption, we recall the asymptotic equipartion property (AEP). The AEP establishes that the output of a random experiment is certain to come from the typical set, under the limit of large number of repetitions. In Ref.~\cite{tomamichel2009} a generalisation of the AEP to quantum theory is developed, establishing that the \textit{smooth min-entropy} $H_{\text{min}}^{\varepsilon}$ (i.e.~the maximum min-entropy for any state $\varepsilon$-close to a fixed state \cite{renner2006}) converges to the conditional von Neumann entropy up to an error term. Our case simplifies this statement to the Shannon entropy, because only classical side information is considered here. Specifically, 
\begin{align}
\frac{1}{N}H_{\text{min}}^{\varepsilon}\left(B^{N}|E^{N},x=2\right) \geq H^{*} - \frac{\delta}{\sqrt{N}} \ ,
\end{align}
for $H_{\text{min}}^{\varepsilon}\left(B^{N}|E^{N},x=2\right)$ being the smooth min-entropy accumulated over a total of $N$ experimental rounds for a particular preparation $x=2$, and $\delta$ a parameter specified in \cite{supp_mat}. By $H^{*}$ we denoted a lower-bound on the single-round Shannon entropy assuming i.i.d.~for a particular set of observable statistics, which can be estimated from \eqref{eq:H_bound}. 

Next, we go beyond the i.i.d.~assumption and perform a second characterization of finite size effects. In this case, we use the entropy accumulation theorem (EAT) generalised to prepare-and-measure scenarios \cite{metger2022,metger2023}, which allows us to quantify the amount of entropy accumulated per round also when they are not i.i.d.. The EAT places a bound on the conditional smooth min-entropy, which depends only on the von Neumann entropy per round plus a correction term which depends on the total number of rounds in the experiment. Every round of the experiment is described by a channel $\mathcal{M}_{i}$ which maps Bob's ($B_{i}$) and Eve's ($E_{i}$) systems to their corresponding systems in the next round ($B_{i+1}$ and $E_{i+1}$ respectively). To make the EAT applicable in our experiment, we assume a no-signaling constraint between Bob and Eve's devices which can be straightforwardly justified thanks to the classical nature of side information. The marginal of the new side information $E_i$ without the new output $B_i$ must be reproducible only with past side information $E_{i-1}$ (i.e. any new $E_i$ should depend on past $E_{i-1}$ alone). This forbids that Eve's information $E_i$ is passed along Bob's systems $B_{i-1}$, which would mean that, in some cases, $E_{i-1}$ holds no information on Bob's outcome but $E_{i}$ does. It follows from assuming classically correlated eavesdroppers that any state Eve and Bob may share is separable and thus, the non-signaling condition is ensured by sealing the laboratory during the prepare-and-measure rounds. In \cite{supp_mat} we explain in more detail how we apply the EAT to our results to certify the randomness outside the i.i.d.\ assumption. Concretely, we are able to bound the smooth-min-entropy on the measurement outcome using the single-round Shannon entropy plus a correction term, i.e.
\begin{align}
\!\frac{1}{N}H_{\text{min}}^{\varepsilon}(\!B^{N}|E^{N}\!\!,x\!=\!2)\!\geq\! \min f_{min}\!(\text{fq.}(b|x)) \!-\! \mathcal{O}(\frac{1}{\sqrt{N}}) ,
\end{align}
where $f_{min}(\text{fq.}(b|x))\leq H^\ast$ is a so-called \textit{min-tradeoff} function, i.e.~an affine function on the observable frequencies $\text{\red{fq.}}(b|x)$ which lower-bounds the minimum Shannon entropy $H^\ast$. This function can be directly taken from the objective function of the dual semidefinite program in \eqref{eq:H_bound}, replacing the ideal probabilities $p(b|x)$ with the observable frequencies, i.e.,
\begin{align}
\!\!\!f_{min}(\text{fq.}(b|x)) = c_m \! - \! \sum_{i=0}^{m-1} \sum_{b,x}\nu_{bx}^{i}\text{fq.}(b|x) + \Tr\left[R^{i}\right] ,
\end{align}
where $\nu_{bx}^{i}$ and $R^{i}$ are the feasible values that maximize the objective function in \eqref{eq:H_bound} (see \cite{supp_mat} for details).

Finally, to extract the randomness from the whole sequence of Bob's outcomes, we use a randomness extractor. Here we shortly introduce the meaning of this function and the main result we use in this work. For a more detailed explanation refer to Ref.~\cite{metger2023}. Consider $H_{\text{min}}^{\varepsilon}(B^{N}|E^{N})\geq h$. A quantum-proof strong $\left(h,\varepsilon_{\text{EXT}}\right)$-extractor is defined as a function EXT that receives a quantum state $\rho_{BE}\otimes\tau_{S}$ and acts on the classical systems $B$ and $S$ with dimension equal to $n_{B} \times n_{S}$ bits. The $l$-bit dimensional output EXT$\left(\rho_{BE}\otimes\tau_{S}\right)$ is $\tilde{\varepsilon}_{\text{EXT}}$-close to $\tau_{L}\otimes\rho_{E}\otimes\tau_{S}$. Here $\tau_L$ and $\tau_S$ are maximally mixed states of dimension $2^{n_S}$ and $2^{l}$, and $\rho_{BE}$ is the classical-quantum state shared by Bob and Eve after the measurement. The input in system $S$ is called the seed of the extractor. The security parameter of this extractor is given by $\tilde{\varepsilon}_{\text{EXT}} := \varepsilon_{\text{EXT}} + 4\varepsilon$, which can be arbitrarily chosen. Furthermore, there exists a quantum-proof strong $\left(h,\varepsilon_{\text{EXT}}\right)$-extractor from which we can extract a total of $l \leq h - 2\log_2\left(1/\varepsilon_{\text{EXT}}\right)$ random bits \cite{renner2006}. In all results from this work we consider $\varepsilon_{\text{EXT}} = 10^{-8}$, which implies an additional correction of $\sim -26/N$ in the extractable randomness per round.

\begin{figure*}
     \centering
     \begin{minipage}[b]{0.49\textwidth}
         \centering
         \includegraphics[trim={0 0 0 0.6cm},clip,width=0.98\textwidth]{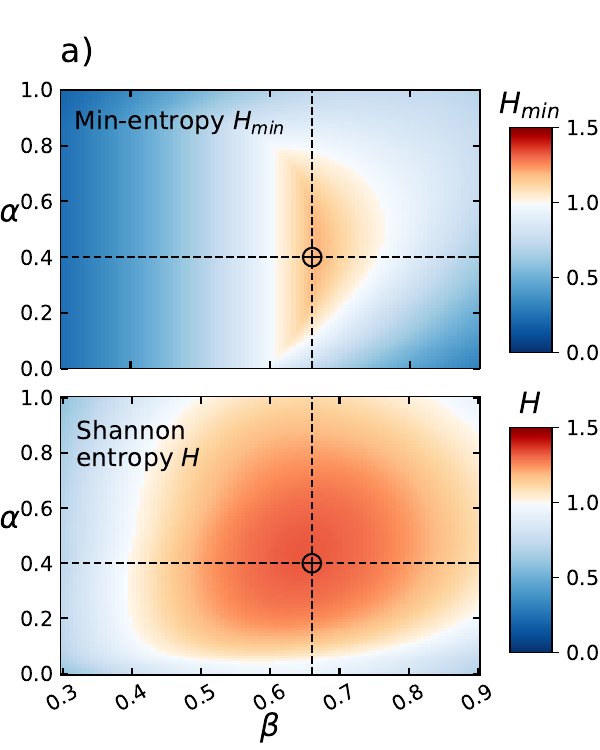}
         \label{fig:cmap}
     \end{minipage}
     \hfill
     \begin{minipage}[b]{0.49\textwidth}
         \centering
         \includegraphics[trim={0 0 1.7cm 0.6cm},clip,width=0.83\textwidth]{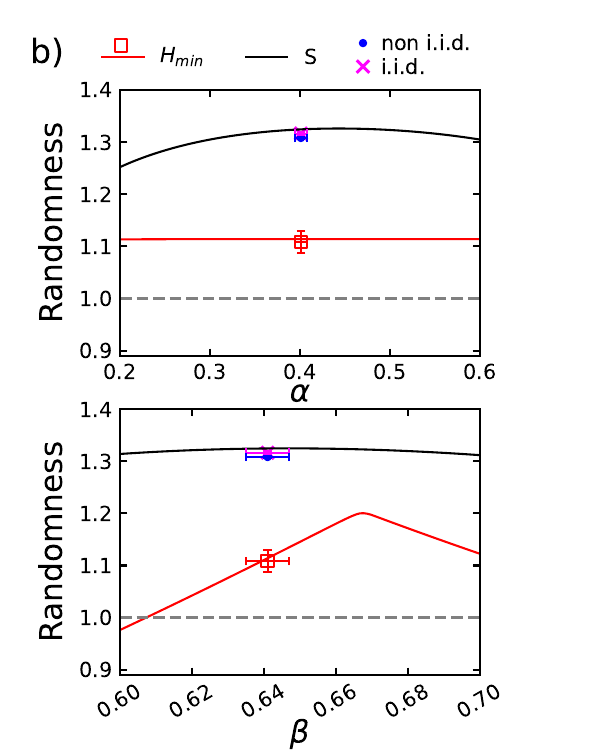}
         \label{fig:experiment}
     \end{minipage}
        \caption{\textbf{a)} Bound on the raw min and Shannon entropies (i.e.~without accounting for finite size effects) as a result from the semidefinite programs presented in \secref{sec:rng} for a limited range of coherent amplitudes. Red regions denote more than one bit of randomness per round. The targeted amplitudes $\alpha_{T}= 0.4$ and $\beta_{T} = 0.66$ are denoted with black-dashed lines. \textbf{b)} Slice of the color plots at the chosen amplitudes. We compare the raw min-entropy, the Shannon entropy and the certifiable randomness through the AEP (i.i.d.)~and generalised EAT (non i.i.d.).~The results of the experiment are shown in a cross/circle/square with their corresponding error bars. The chosen parameters are $m=8$, $\varepsilon=10^{-8}$ with photon loss $1-\eta=0.06$, dark count probability of $p_{\text{dc}}=10^{-6}$ and total number of samples $N=10^{7}$ over $11$ distinct repetitions.}
        \label{fig:cmap_experiment}
\end{figure*}

\subsection{Implementation}

We implement the protocol using time-bin encoded coherent states and single-photon detection. Our approach is inspired by previous works in similar settings \cite{brask2017,tebyanian2021}. The setup is illustrated in \figref{fig:implementation}. In the following, we detail its working principle. 

Our photon source is a 1550.32 nm continuous-wave laser. We use an electro-optic modulator (EOM) and two acousto-optic modulators (AOMs) to carve the output of the laser into 10 ns-width pulsed coherent states with appropriate amplitude.  We alternate the experiment setup between a locking phase and a randomness-generation stage. During the randomness-generation stage, the two AOMs together induce a stable attenuation of 77 dB to bring the laser power down to single-photon level. This attenuation is removed during the locking stage to monitor the transmission power of the EOM and correctly set up its bias voltage. In doing so, we achieve a dark count rate of only $p_{\text{dc}} = (4.5 \pm 2.8) \times 10^{-6}$ per pulse. The EOM, on the other hand, generates the correct amplitudes for the input states during a 10 ns time window, which is chosen as the duration of the time-bin. In each round, the pulses can emerge in an early or a late time-bin. The two self-testing states for $x=0$ and $x=1$ are prepared using a coherent state with amplitude $\alpha$ in the early and late bins, respectively. We write this as $\ket{\psi_0}=\ket{\alpha 0}$ and $\ket{\psi_1}=\ket{0 \alpha}$. On the other hand, the third state, $x=2$, used for randomness extraction is prepared using amplitude $\beta$ in both early and late bins, $\ket{\tilde{\psi}_2}=\ket{\beta \beta}$. The early and late pulses are separated by 800 ns and the full protocol runs at a 0.526 MHz rate. To ensure high efficiency of the photodetection, we optimize the polarization state of the photons with a fiber polarization controller before the superconducting detector and splice all the fiber connection points in the measurement stage to minimize the photon loss.

The measurement is performed by a superconducting single-photon detector (ID Quantique ID281) with a quantum efficiency of $94\%$ which detects photons from the carved, attenuated laser beam. The values for $\alpha$ and $\beta$ are calibrated calculating the click rate related to the coherent sates. Further, we have introduced a feedback control loop to stabilize the output power of the EOM to stabilize the coherent state amplitudes throughout the experiment. From the measured counting rates, we deduce the coherent state amplitude we sent to be $\alpha_{L} = 0.401 \pm 0.007$ and $\beta_{L} = 0.641 \pm 0.006$. Here, the $1\sigma$ standard deviations are caused by the residual intensity fluctuations across the data collection time of the entire experiment.

The outcome $b$ is determined by in which time bin the detector clicks. We label events with clicks in only the early time-bin by $b=0$, events with clicks only in the late time-bin by $b=1$, events with no clicks in either bin by $b=2$, and events with clicks in both bins by $b=3$. The fourth outcome does not trigger during the self-test rounds and thus, is not relevant in our framework. Therefore, as discussed in \cite{supp_mat}, the outcome $b=3$ is binned with the first two, leaving the measurement effectively with three outcomes. Using the measurement setup, in a calibration step, we also determine the AOM and EOM voltages required to prepare coherent states $\ket{\alpha}$ and $\ket{\beta}$ with the desired overlap.

In order to reduce finite size effects, we take around $10^7$ rounds of measurements over $11$ distinct iterations. After gathering every $10^7$ round of data on each iteration, we perform a calibration measurement before taking the data again. The fluctuation of the count rate, measured by the standard deviation across all experiment runs, is only 1.52\% relative to the average value, certifying the consistency between the states prepared at different times. \\

\section{Results} \label{sec:results}

In order to find optimal experimental settings, we first simulate the system, computing the randomness for varying coherent-state amplitudes. In \figref{fig:cmap_experiment} we show the min-entropy under realistic conditions (including loss and detector dark counts). The results show maximal min-entropy around the amplitudes $\alpha_{T}=0.4$ and $\beta_{T}=0.66$, which we therefore target in the experiment. This corresponds to states $\ket{\psi_x}$ with overlaps $|\braket{\psi_0|\psi_1}| = 0.84$ and $|\braket{\psi_0|\tilde{\psi}_2}|=|\braket{\psi_1|\tilde{\psi}_2}| = 0.78$. We note these fulfil the criterion $|\braket{\psi_0|\psi_1}|\geq 1/5$ deduced from \eqref{eq:varphi_USD} for the possibility of equiprobable outcomes. These preparations should thus allow correlations yielding high values of randomness. We additionally run the semidefinite program optimisation method to efficiently compute direct bounds on the Shannon entropy using the exact same realistic settings. Comparing with the min-entropy results, we find a marked increase in the certifiable randomness, as well as a vast expansion of the usable parameter region. Although the optimal randomness in this case is not exactly given by the targeted amplitudes $\alpha_{T}$ and $\beta_{T}$, we keep them to directly compare the best possible certifiable randomness using the min-entropy vs.~the Shannon entropy. 

The overlaps constrain support outside the span of $\ket{\psi_0}$ and  $\ket{\psi_1}$ but do not eliminate it. Specifically, the minimal amplitude introduced in \eqref{eq:third_tilde} is $a \geq 0.66$ on the targeted coherent amplitudes. Moreover, the conditional probability $p(b|2)$ does not depend on $\alpha$. However, we do see some dependency on $\alpha$ in the randomness in \figref{fig:cmap_experiment}, which becomes more evident for higher values of $\beta$. The fact that $a<1$ for our amplitudes is the main responsible of that dependency and also implies that the correlations are not reproducible only by unique and extremal POVMs. This reduces the secrecy of the outcome, decreasing the certifiable randomness. Nevertheless, we are able to find a set of amplitudes where, although Eve has unbounded dimensionality, her guessing probability is lower than $1/2$, i.e.\ more than one bit of randomness is certified using the min-entropy and even more with the Shannon entropy. 

\begin{figure}
    \centering
    \includegraphics[width=0.47\textwidth]{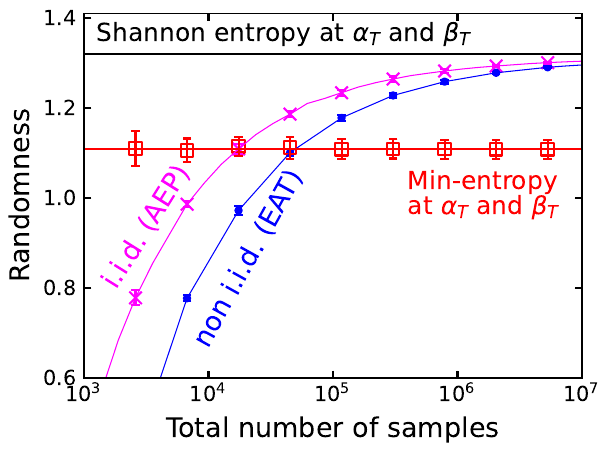}
    \caption{Scaling of i.i.d.\ and non i.i.d.\ bounds with the number of rounds in the experiment. In our experiment we took $\sim\!10^7$ samples, which explains the almost-equivalency between both approaches in our results.}
    \label{fig:scalability}
\end{figure}

We run the experiment through the proposed implementation and evaluate the randomness for a single configuration of targeted amplitudes. We show the obtained results in \figref{fig:cmap_experiment} (right-side plots). The experimental data we collected indicate that the actual amplitudes used in the lab were $\alpha_{L}=0.401 \pm 0.007$ and $\beta_{L}=0.641 \pm 0.006$. Our observations agree well with the predictions from the simulations, 
and show a randomness extraction
 rate of $1.304\pm 0.002$ and $1.299\pm 0.002$ bits per round after $10^7$ runs of the experiment, when analysed assuming i.i.d.~experiment rounds using the asymptotic equipartition property (AEP) and without i.i.d.~rounds using the entropy accumulation theorem (EAT), respectively. Additionally, we compute the raw min-entropy per round without taking into account finite size effects and, using the experimental data, we obtain $1.11\pm 0.02$ bits of randomness per round. All error bars (also shown in \figref{fig:cmap_experiment}) are the standard deviation obtained over running the experiment $11$ different times. We observe approx.~$19 \%$ improvement in the certifiable randomness by using the Shannon entropy (which is computed in the AEP and EAT approaches) compared to the min-entropy. Also, the estimated errors using the AEP and EAT approaches are lower than the min-entropy approach by one order of magnitude. This is mainly because the Shannon entropy is more stable than the min-entropy against fluctuations on the coherent amplitudes. Furthermore, the reason behind the almost-equivalence between the i.i.d.\ and non i.i.d.~value is due to the high amount of samples gathered in the experiment (around $10^7$) which foresees the convergence of the EAT to the AEP in the limit of infinite number of experimental rounds. In \figref{fig:scalability} we additionally collected random subsets between $10^3$ and $10^7$ data-points form the $11$ different gathered experimental data-sets and computed the randomness using the raw min-entropy, the i.i.d.~(AEP) and non i.i.d.~(EAT) approaches. Observe that the total of $10^{7}$ experimental rounds lies close to the asymptotic limit. Nevertheless, note that the non i.i.d.~approach is still able to certify more randomness than the raw min-entropy down to approximately $3.8\times 10^{4}$ samples.

\section{Conclusion}
\label{sec.discussion}

We have realized a semi-DI randomness generation protocol implemented in an optical platform, as shown in \figref{fig:implementation}. By exploiting the entropy accumulation theorem generalised to prepare-and-measure scenarios, we bound the certifiable randomness in terms of the single-round Shannon entropy without assuming that experimental rounds are independent and identically distributed (i.i.d.) and including finite-size effects. We proposed a method for bounding the Shannon entropy in our semi-DI framework based on overlap assumptions. This consists in expanding the logarithm function on its integral form using the Gauss-Radau quadrature, i.e.\ employing the methods from \cite{brown2023} to derive bounds on the von Neumann entropy and imposing classical side information. We show that this allows certifying more randomness than a standard approach bounding the single-round min-entropy under an i.i.d.\ assumption and neglecting finite-size effects. Thus, we certify more randomness under much less restrictive assumptions.

\acknowledgements

C.R.C.~is supported  by the  Wenner-Gren Foundation. We gratefully acknowledge support from the Danish National Research Foundation, Center for Macroscopic Quantum States (bigQ, DNRF142), VILLUM FONDEN (research grant 40864), the Carlsberg Foundation CF19-0313, Innovation Fund Denmark (PhotoQ project, grant no. 1063-00046A), the Horizon Europe MSCA (\href{https://doi.org/10.3030/101106833}{Project GTGBS}, grant no. 101106833) and a DTU-KAIST Alliance Stipend. \\

%

\section*{Code availability}

The codes used to generate the results in this paper are available in GitHub: \url{https://github.com/chalswater/QRNG_entropy_accumulation}. Experimental data is available upon request.

\bibliography{main}

\begin{widetext}

\appendix

\section{Min-entropy semidefinite program}
\label{app.sdp}

In this section we formally introduce the semidefinite program we use to bound the certifiable randomness using the min-entropy. We start presenting the primal form of the problem. Our goal is to maximise the guessing probability of the eavesdropper which we can write as
\begin{align}
    p_g = \sum_{\lambda}q(\lambda)\underset{b}{\max}\left\{\Tr\left[\rho_2 \pi_{b}^{\lambda}\right]\right\} \ .
\end{align}
The maximisation is done through all possible measurement strategies $\lambda$, distributions $q(\lambda)$ and POVM elements $\pi_{b}^{\lambda}$. These are constrained to be valid distributions and POVMs, which implies
\begin{align}
    & q(\lambda) \geq 0  && \sum_{\lambda}q(\lambda)=1 && q(\lambda)\in\mathds{R} \\
    & \pi_{b}^{\lambda}\succeq 0 && \sum_{b}\pi_{b}^{\lambda}= \mathds{1} && \pi_{b}^{\lambda}=\left(\pi_{b}^{\lambda}\right)^{\dagger} \ .
\end{align}
There is the additional constraint that the observed probabilities must be reproduced on the real experiment. This is reflected in
\begin{align}
    p(b|x) = \sum_{\lambda}q(\lambda)\Tr\left[\rho_x\pi_{b}^{\lambda}\right] \ .
\end{align}
Since states $\rho_x$ are not fully specified, but instead only their overlaps are bounded, we will insert the states in \eqref{eq:tilde_states}, so $\rho_x=\ket{\tilde{\psi}_x}\bra{\tilde{\psi}_x}$.

This optimisation problem can be rendered as a linear semi-definite program following a couple of steps. First, we will consider only the most relevant steategies, which in our case are those which yield the maximal value $\underset{b}{\max}\left\{\Tr\left[\rho_x \pi_{b}^{\lambda}\right]\right\}$. This can be done by simply labeling $\lambda=b$ the maximal strategy for outcome $b$, i.e. $\underset{b}{\max}\left\{p(b|x=2,\lambda)\right\}=p(\lambda|x=2,\lambda)$. This leaves us with only $n_{\text{B}}$ relevant strategies, being $n_{\text{B}}$ the number of different outcomes from the measurement. Secondly, we will absorb the distribution $q(\lambda)$ in the POVM element $\pi_{b}^{\lambda}$ and define a new quantity $M_{b}^{\lambda}=q(\lambda)\pi_{b}^{\lambda}$. The definition of this new operator changes the above constraints to the following:
\begin{align}
    M_{b}^{\lambda} \succeq 0 \quad , \qquad M_{b}^{\lambda}=\left(M_{b}^{\lambda}\right)^{\dagger} \ \forall b,\lambda \quad , \qquad \sum_{b}M_{b}^{\lambda} = \frac{1}{D}\Tr\left[\sum_{b}M_{b}^{\lambda}\right] \ \forall\lambda  \ ,
\end{align}
where $D$ is the dimensionality of the eavesdropper. The useful space accessible by the eavesdropper is that spanned by the states involved in the experiment. Since we are considering a three-state discrimination setting, the dimension can be at maximum the number of states, i.e. $D=3$. The reproducibility constraint is also changed to simply
\begin{align}
    p(b|x) = \sum_{\lambda}\Tr\left[\rho_{x}M_{b}^{\lambda}\right]  \ .
\end{align}
Finally, we can re-write the guessing probability in the following way
\begin{align}
    p_{g} = \sum_{\lambda}\Tr\left[\rho_{2}M_{\lambda}^{\lambda}\right] \ .
\end{align}
An upper bound $p_{g}^{*}\geq p_{g}$ can be found by maximising it through all possible $2\times 2$ matrices $M_{b}^{\lambda}$ that fulfil the constraints above.

\section{Shannon entropy semidefinite program}\label{ap:seesaw}

In this section we present method we used to bound the Shannon entropy and the semidefinite program to numerically compute it. We begin presenting the primal formulation of the problem, for then deriving the dual version we use to define the min-tradeoff function used in the main text.

\subsection{Primal semidefinite program}

Consider a prepare-and-measure scenario where Alice draws a classical value $x$ with probability $p_x$ and prepares a pure quantum sate $\rho_x$. This state is sent to Bob, who will perform a measurement $\pi_b$ and spit $b$ as a measurement outcome.

Consider the prepare-and-measure scenario where Alice draws a classical value $x$ with probability $p_x$ which she feeds into a device that prepares a quantum state $\Psi_{x}$. While she sends one part of $\Psi_{x}$ (namely $\rho_x$) to Bob, the rest (namely $\sigma_x$) is leaked to the adversary (Eve). Let us label Bob and Eve's Hilbert spaces with $\mathcal{H}_{B}$ and $\mathcal{H}_{E}$ respectively, such that the global state is $\Psi_{x}\in\mathcal{H}_{B}\otimes\mathcal{H}_{E}$, $\rho_x:=\Tr_{E}\left[\Psi_x\right]$ and $\sigma_x:=\Tr_{B}\left[\Psi_x\right]$. Then, Bob performs a measurement on $\rho_x$ in his laboratory, while Eve's goal is to guess Bob's measurement outcome by measuring her part of the state $\sigma_x$ on her separated laboratory, and also be able to share randomness (through the hidden variable $\lambda$) with Bob's measurement device. Let us call $\left\{\pi_b^\lambda\right\}$ the POVM that represents the actual measurement performed in Bob's device.

Our first goal is to find an expression to lower-bound the von Neumann entropy in Bob's measurement outcome relative to Eve. After Bob measures, the whole quantum state shared between Bob and Eve is updated to the classical-quantum state
\begin{align}
    \tilde{\Psi}_{x} = \sum_{b}\ket{b}\bra{b}\otimes \tilde{\sigma}_{b|x} \ .
\end{align}
This means that Bob is left with the classical value $b$ of his measurement outcome, while Eve's state becomes $\tilde{\sigma}_{b|x} = \sum_\lambda q_\lambda \Tr_{B}\left[\left(\pi_{b}^\lambda\otimes\mathds{1}\right)\Psi_x^{\lambda}\right]$. We can express the von Neumann entropy of Bob's outcome conditioned on Eve's knowledge for a particular state preparation $x^{*}$ in terms of the relative entropy
\begin{align}
    S\left(B|E,X=x^{*}\right)=-D\left(\tilde{\Psi}_{x^{*}}||\mathds{1}\otimes \bar{\sigma}_{x^*}\right) \ ,
\end{align}
for $\bar{\sigma}_x=\sum_b \tilde{\sigma}_{b|x}$. As shown in Ref.~\cite{brown2023}, applying the Gauss-Radau quadrature to an integral representation of the logarithm yields a variational upper bound for the quantum relative entropy,
\begin{align}
    D\left(\rho||\sigma\right) \leq \sum_{i=1}^{m-1}\frac{w_{i}}{t_{i}\log{2}}\underset{Z_i}{\inf}\left(1+\Tr\left[\rho\left(Z_i + Z_i^{\dagger}+(1-t_i)Z_i^{\dagger}Z_i\right)\right]+t_i\Tr\left[\sigma Z_i Z_i^{\dagger}\right]\right) \ ,
\end{align}
for $Z_i$ being arbitrary complex matrices and $\left\{w_i,t_i\right\}$ the weights and nodes from the Gauss-Radau quadrature \cite{bouzitat1952,walter1991}, respectively. In our case, this turns into the following variational lower bound on the von Neumann entropy for a particular preparation $x^{*}$,
\begin{align}
\label{eq:hbe_bound}
    &S\left(B|E,X=x^{*}\right)\geq \\
    &c_m + \sum_{i=1}^{m-1}\tau_i\sum_{b,\lambda}q(\lambda)\underset{\left\{Z^{b}_{i},F^{b}_{i},K^{b}_{i}\right\}}{\inf}\Tr\left[\Psi_{x^*}^{\lambda}\left(\pi_{b}^{\lambda}\otimes \left(Z_{i}^{b} + Z_{i}^{b\dagger}+(1-t_i)F_{i}^{b}\right)\right) + t_{i}\Psi_{x^*}\left(\mathds{1}_{B}\otimes K_{i}^{b}\right)\right]  \nonumber
\end{align}
where we defined $\tau:=\frac{w_{i}}{t_{i}\log{2}}$ $c_{m}:=\sum_{i=1}^{m-1}\tau_i$ together with the matrices
\begin{align}
    Z_{i}^{b} :=& \Tr_{B}\left[\left(\ket{b}\bra{b}\otimes\mathds{1}\right)Z_i\right] \\
    F_{i}^{b} :=& \Tr_{B}\left[\left(\ket{b}\bra{b}\otimes\mathds{1}\right)Z^{\dagger}_{i}Z_i\right]  \\
    K_{i}^{b} :=& \Tr_{B}\left[\left(\ket{b}\bra{b}\otimes\mathds{1}\right)Z_{i}Z_{i}^{\dagger}\right]  \ .
\end{align}

In our scenario we confider that the eavesdropper has only access to side classical information. We label with $\lambda$ the collection of physical parameters known by Eve which determine the behaviour of Bob's device. This infers that the measurement performed in Bob's device consists of a convex mixture of measurements, determined by the hidden variable $\lambda$ distributed according to $q(\lambda)$ and publicly announced to Eve. In this case, after Bob performs his measurement, Eve's part of the state updates to $\tilde{\sigma}_{b|x} = \sum_{\lambda}q(\lambda)\Tr\left[\rho_{x}\pi_{b}^{\lambda}\right]\ket{\lambda}\bra{\lambda}$. This translates the previous bound on the von Numann entropy \eqref{eq:hbe_bound} into the following Shannon entropy bound,
\begin{align}
\label{eq:vn_cc}
    H\left(B|E,X=x^{*}\right)\geq c_m + \sum_{i=1}^{m-1}\frac{w_{i}}{t_{i}\log{2}} \ \underset{z^{\lambda b}_{i},\eta^{\lambda b}_{i}}{\inf} \ \sum_{b,\lambda}q(\lambda)\Tr\left[\pi^{\lambda}_{b}\rho_{x^{*}}\right]\left(z_{i}^{\lambda b} + z_{i}^{\lambda b *}+(1-t_i)\eta_{i}^{\lambda b}\right)+q(\lambda)t_i \eta_{i}^{\lambda b} \ ,
\end{align}
where we introduced the following scalar variables
\begin{align}
\label{eq:hbe_class}
    z_{i}^{\lambda b} :=& \Tr\left[\left(\ket{b}\bra{b}\otimes \ket{\lambda}\bra{\lambda}\right)Z_i\right] \\
    \eta_{i}^{\lambda b} :=& \Tr\left[\left(\ket{b}\bra{b}\otimes \ket{\lambda}\bra{\lambda}\right)Z^{\dagger}_{i}Z_i\right] = \Tr\left[\left(\ket{b}\bra{b}\otimes \ket{\lambda}\bra{\lambda}\right)Z_{i}Z_i^{\dagger}\right] \ ,
\end{align}
defined from projecting the arbitrary matrices $Z_i$ onto $\ket{b}\bra{b}$ and $\ket{\lambda}\bra{\lambda}$ in Bob and Eve's laboratories, respectively. Observe that these definitions imply that all scalars $z_{i}^{\lambda b}$ and $\eta_{i}^{\lambda b}$ are constrained to fulfil
\begin{align}
\label{eq:ct_psd}
    \begin{pmatrix}
        1 & z_{i}^{\lambda b} \\
        \left(z_{i}^{\lambda b}\right)^{*} & \eta_{i}^{\lambda b}
    \end{pmatrix} \succeq 0 \ .
\end{align}
One can see that by constructing the completely positive channel $\Lambda_{b,\lambda}\left[M\right]:=\Tr\left[\left(\ket{b}\bra{b}\otimes \ket{\lambda}\bra{\lambda}\right)M\right]$, which consists of a projection onto classical outputs $b$ and $\lambda$. Applying this channel though each element of the strictly positive-semidefinite matrix $P_i:=\ket{\tau_i}\bra{\tau_i}$ built from the inner product of vectors $\ket{\tau_i}:=\left(\mathds{1},Z_{i}\right)$ yields the matrix in \eqref{eq:ct_psd}, which by definition must be positive-semidefinite. Furthermore, we can restrict to $z_{i}^{\lambda b}$ as real scalar parameters, as adding a complex phase in the infimum from \eqref{eq:vn_cc} involves a trivial component with no effects on the optimal solution.

In order to find a lower bound on the Shannon entropy, we need to minimise the right-hand side of \eqref{eq:vn_cc}. We aim to find a linear convex problem in order to formulate it as a semidefinite program. To do so, we first define the following matrices
\begin{align}
	M_{b} :=& \sum_\lambda q(\lambda) \pi_{b}^\lambda \label{eq:M} \\
	\xi_{b}^{ia} :=& \sum_\lambda q(\lambda) z_i^{a\lambda} \pi_{b}^\lambda \label{eq:xi} \\
	\theta_{b}^{ia} :=& \sum_\lambda q(\lambda) \eta_i^{a\lambda} \pi_{b}^\lambda \label{eq:theta} \ .
\end{align}
Now let us define the following block matrix,
\begin{align}
	\mathcal{G}_{b}^{ia}:=\begin{pmatrix}
        M_{b} & \xi_{b}^{ia} \\
        \xi_{b}^{ia} & \theta_{b}^{ia}
    \end{pmatrix} = \sum_\lambda q_\lambda \pi_b^\lambda \otimes \begin{pmatrix}
        1 & z_{i}^{\lambda b} \\
        z_{i}^{\lambda b} & \eta_{i}^{\lambda b}
    \end{pmatrix}  \ .
\end{align}
By construction, since $\pi_b^\lambda \succeq 0$ and \eqref{eq:ct_psd}, $\mathcal{G}_{b}^{ia}\succeq 0$. The reverse condition is not necessarily true and thus, any optimisation over the matrices \eqref{eq:M}, \eqref{eq:xi} and \eqref{eq:theta} constrained by $\mathcal{G}_{b}^{ia}\succeq 0$ implies a semidefinite relaxation of the original problem. Finally, we need to impose normalisation of $M_b$, constrain the observable correlations $p(b|x)$ and normalisation of $\xi_b^{ia}$ and $\theta_{b}^{ia}$. Once we put all together, the final optimisation reads as follows
\begin{align}
    \underset{\left\{M_{b},\xi_{b}^{ia},\theta_{b}^{ia}\right\}}{\text{minimize}} & \quad c_m + \sum_{i=1}^{m-1} \tau_i \sum_{a}  \Tr\left[\rho_{x^{*}}\left(2\xi_{a}^{ia}+(1-t_i)\theta_{a}^{ia} + t_i \sum_{b}\theta_{b}^{ia}\right)\right]  \label{eq:sdp_relax_1} \\    
    \text{subject to} \qquad & \quad \begin{pmatrix}
        M_{b} & \xi_{b}^{ia} \\
        \xi_{b}^{ia} & \theta_{b}^{ia}
    \end{pmatrix} \succeq 0 , \quad \sum_{b}M_{b} = \mathds{1} , \quad \Tr\left[M_{b} \rho_x\right] = p(b|x) ,  \nonumber \\
    & \quad \sum_{b}\xi_{b}^{ia} = \frac{1}{D}\Tr\left[\sum_{b}\xi_{b}^{ia}\right]\mathds{1} , \quad \sum_{b}\theta_{b}^{ia} = \frac{1}{D}\Tr\left[\sum_{b}\theta_{b}^{ia}\right]\mathds{1}  \nonumber 
\end{align}
Finally, note that the indexing $i$ does not play any relevant role in the optimisation. To reduce computation costs, we can thus perform a simplification and run the optimisation independently for each $i$ and then sum up the result. That is,

\begin{align}
    H(B|E,X=x^\ast) \geq c_m + \sum_{i=1}^{m-1} \underset{\left\{M_{b},\xi_{b}^{a},\theta_{b}^{a}\right\}}{\text{minimize}} & \quad \sum_{a}  \tau_i\Tr\left[\rho_{x^{*}}\left(2\xi_{a}^{a}+(1-t_i)\theta_{a}^{a} + t_i \sum_{b}\theta_{b}^{a}\right)\right]  \label{eq:sdp_relax} \\    
    \text{subject to} \ & \quad \begin{pmatrix}
        M_{b} & \xi_{b}^{a} \\
        \xi_{b}^{a} & \theta_{b}^{a}
    \end{pmatrix} \succeq 0 , \quad \sum_{b}M_{b} = \mathds{1} , \quad \Tr\left[M_{b} \rho_x\right] = p(b|x) ,  \nonumber \\
    & \quad \sum_{b}\xi_{b}^{a} = \frac{1}{D}\Tr\left[\sum_{b}\xi_{b}^{a}\right]\mathds{1} , \quad \sum_{b}\theta_{b}^{a} = \frac{1}{D}\Tr\left[\sum_{b}\theta_{b}^{a}\right]\mathds{1}  \nonumber 
\end{align}

Note that now we removed the $i$ indexing from all variables which are reset after each independent optimisation. Any feasible solution that satisfies the constraints in \eqref{eq:sdp_relax} provides a lower bound on the Shannon entropy in \eqref{eq:vn_cc}.

\subsection{Dual semidefinite program}

We now derive the dual formulation of the semidefinite program above. We begin writing the Lagrangian of the problem, which reads $\mathcal{L} = c_m + \sum_{i=0}^{m-1}\mathcal{L}_i$ for
\begin{align}
\mathcal{L}_i =\ & \sum_{a} \tau_i \Tr\left[\rho_{x^{*}}\left(2\xi_{a}^{a}+(1-t_i)\theta_{a}^{a} + t_i \sum_{b}\theta_{b}^{a}\right)\right] \\
-& \sum_{a,b}\Tr\left[M_{b}E^{a}_{b} + \xi_{b}^{a}\left(F^{a}_{b}+F^{a\dagger}_{b}\right) + \theta_{b}^{a}L^{a}_{b}\right] + \sum_{b,x}v_{b,x}\left(\Tr\left[M_{b} \rho_x\right] - p(b|x)\right) \nonumber \\
+& \sum_{a}\Tr\left[Q^{a}\left(\sum_{b}\xi_{b}^{a} - \frac{1}{d}\Tr\left[\sum_{b}\xi_{b}^{a}\right]\mathds{1}\right)\right] + \sum_{a}\Tr\left[P^{a}\left(\sum_{b}\theta_{b}^{a} - \frac{1}{d}\Tr\left[\sum_{b}\theta_{b}^{a}\right]\mathds{1}\right)\right] \\
+&\Tr\left[R\left(\sum_{b}M_{b} - \mathds{1}\right)\right] \nonumber \ . \\
\end{align}
Here the we introduced the $D\times D$ matrices $E^{a}_{b}$, $F^{a}_{b}$ and $L^{a}_{b}$ that correspond to the blocks of the positive semidefinite matrix
\begin{align}
\begin{pmatrix}
E^{a}_{b} & F^{a}_{b} \\
F^{a\dagger}_{b} & L^{a}_{b}
\end{pmatrix} \succeq 0 \ , 
\end{align}
which corresponds to the dual variable of the positive-semidefinite constraint from the primal semidefinite program. We also added the scalar dual variables $\nu_{bx}$ corresponding to the primal constraint restricting the observable probabilities $p(b|x)$ and the $D\times D$ dual variables $R$, $Q^{a}$ and $P^{a}$ for the normalisation primal constraints.

The solution to the original semidefinite program is always a saddle point of the Lagrangian $\mathcal{L}$, identified among the stationary points. The infimum of the Lagrangian over the variables of the primal semidefinite program reads
\begin{align}
\mathcal{I} &= \underset{\left\{M_{b},\xi_{b}^{a},\theta_{b}^{a}\right\}}{\inf} \mathcal{L} = c_m + \sum_{i=1}^{m-1} \underset{\left\{M_{b},\xi_{b}^{a},\theta_{b}^{a}\right\}}{\inf} \mathcal{L}_i \\
&= c_m + \sum_{i=1}^{m-1}\left\{-\sum_{b,x}v_{b,x}p(b|x) - \Tr\left[R\right] + \sum_b \Tr\left[M_b A_b\right] + \sum_{a,b}\Tr\left[\xi_b^{a}X_b^{a}\right]+\sum_{a,b}\Tr\left[\theta_b^{a}T_b^{a}\right]\right\} \nonumber
\end{align}
where $A_b$, $X_b^{a}$ and $T_b^{a}$ collect all terms in $\mathcal{L}_i$ multiplying the primal variables $M_b$, $\xi_b^{a}$ and $\theta_b^{a}$ respectively. The infimum $\mathcal{I}$ represents a lower bound on the
optimal solution of the primal semidefinite program. However, note that the infimum will diverge unless $A_b$, $X_b^{a}$ and $T_b^{a}$ are nullified. To
get good bounds therefore, we must maximise $\mathcal{I}$ over the set of dual variables that satisfy $A_b=0$, $X_b^{a}=0$ and $T_b^{a}=0$. This maximisation
leads to the dual formulation of the semidefinite program,
\begin{align}
    H(B|E,X=x^\ast)\geq c_m + \sum_{i=1}^{m-1} \ \underset{\begin{pmatrix}
    E^{a}_{b},F^{a}_{b},L^{a}_{b} \\
    R,\nu_{bx},Q^{a},P^{a}
    \end{pmatrix}
    }{\text{maximize}} & \quad - \sum_{b,x}\nu_{b,x}p(b|x) - \Tr\left[R\right] \label{eq:dual_sdp} \\    
    \text{subject to} & \quad \begin{pmatrix}
        E_{b}^{a} & F_{b}^{a} \\
        F_{b}^{a\dagger} & L_{b}^{a}
    \end{pmatrix} \succeq 0  \nonumber \\
    & \quad \sum_{a} E_{b}^{a} = \sum_{x}\rho_x \nu_{bx} 
     + R \nonumber   \\
& \quad F_{b}^{a} + F_{b}^{a\dagger} = 2\tau_i\rho_{x^\ast}\delta_{a,b} + Q^{a} - \frac{1}{d}\Tr\left[Q^{a}\right]\mathds{1} \nonumber \\
& \quad L_{b}^{a}\!=\!\tau_i\rho_{x^\ast}\!\left[\left(1\!-\!t_i\right)\delta_{a,b}\!+\!t_i\right]\! +\! P^{a}\! -\! \frac{1}{d}\Tr\left[P^{a}\right]\mathds{1}   \nonumber \ .
\end{align}
Any set of $D\times D$ matrices $E_{b}^{a}$, $F_{b}^{a}$, $L_{b}^{a}$, $R$, $Q^{a}$, $P^{a}$ and scalars $\nu_{bx}$ that satisfy the above constraints represent a valid lower bound on the Shannon entropy.

\section{Unconstrained dimensionality and semi-device independence}
\label{ap:olaps}

In this section, we show how bounding the overlaps between the prepared states can limit this the use of any additional dimension by the eavesdropper.

Consider the state discrimination scenario with three preparations. Two of the prepared states $\ket{\tilde{\psi}_{0}}$ and $\ket{\tilde{\psi}_{1}}$ have support only on a two-dimensional Hilbert space, but the third state $\ket{\tilde{\psi}_{2}}$ may have support also on a third dimension. Then, 
\begin{align}
    &\ket{\tilde{\psi}_{0}} = \cos\frac{\phi}{2}\ket{0} + \sin\frac{\phi}{2}\ket{1} + 0 \ket{2} \\
    &\ket{\tilde{\psi}_{1}} = \cos\frac{\phi}{2}\ket{0} - \sin\frac{\phi}{2}\ket{1} + 0\ket{2} \\
    &\ket{\tilde{\psi}_{2}} = \sqrt{a}\left(\cos\frac{\theta}{2}\ket{0} + e^{i\varphi}\sin\frac{\theta}{2} \ket{1}\right) + \sqrt{1-a}\ket{2} 
\end{align}
Suppose now that we only trust a bound on the overlap of the prepared states. Let us define $|\braket{\tilde{\psi}_{0}|\tilde{\psi}_{1}}|\geq|d_{01}|$, $|\braket{\tilde{\psi}_{0}|\tilde{\psi}_{2}}|\geq|d_{02}|$ and $|\braket{\tilde{\psi}_{1}|\tilde{\psi}_{2}}|\geq|d_{12}|$. 
This means that 
\begin{align}
\label{eq:bound_on_a}
    \left|\braket{\tilde{\psi}_0|\tilde{\psi}_2}\right|^{2} =& a\left(\cos\frac{\phi}{2}\cos\frac{\theta}{2} + e^{i\varphi}\sin\frac{\phi}{2}\sin\frac{\theta}{2}\right)\left(\cos\frac{\phi}{2}\cos\frac{\theta}{2} + e^{-i\varphi}\sin\frac{\phi}{2}\sin\frac{\theta}{2}\right) \\
    =& a\left(\cos^{2}\frac{\phi}{2}\cos^{2}\frac{\theta}{2} + \sin^{2}\frac{\phi}{2}\sin^{2}\frac{\theta}{2} + \left(e^{i\varphi}+e^{-i\varphi}\right)\cos\frac{\phi}{2}\cos\frac{\theta}{2}\sin\frac{\phi}{2}\sin\frac{\theta}{2}\right) \\
    =& \frac{a}{2}\left(1+\cos\phi\cos\theta + \sin\varphi\sin\phi\sin\theta\right) \\
    \left|\braket{\tilde{\psi}_1|\tilde{\psi}_2}\right|^{2} =& \ldots = \frac{a}{2}\left(1+\cos\phi\cos\theta - \sin\varphi\sin\phi\sin\theta\right) \\
    \left|\braket{\tilde{\psi}_0|\tilde{\psi}_2}\right|^{2} + \left|\braket{\tilde{\psi}_1|\tilde{\psi}_2}\right|^{2} =& a\left(1+\cos\phi\cos\theta\right) \geq |d_{02}|^{2}+|d_{12}|^{2} \\
    a\geq&\frac{d_{02}^{2}+d_{12}^{2}}{1+d_{01}\cos\theta}
\end{align} 
If $a=1$, the three states will have support on the same bi-dimensional Hilbert space. Whilst the measurement device is treated as a black box, the preparation device is partially characterized through the bounds we place on the overlaps of the prepared states. Then, the eavesdropper has the freedom in choosing the states $\ket{\psi_{x}}$ in terms of the angles $\phi$, $\theta$ and $\varphi$ that satisfy those bounds. Her probability of guessing the measurement outcome when state $\ket{\tilde{\psi}_2}$ is prepared is
\begin{align}
    p_g &= \sum_{\lambda}q(\lambda)\underset{b}{\max}\left\{\bra{\tilde{\psi}_{2}}\pi_{b}^{\lambda}\ket{\tilde{\psi}_{2}}\right\} \\
    &= \sum_{\lambda}q(\lambda)\underset{b}{\max}\left\{a\bra{\psi_{2}}\pi_{b}^{\lambda}\ket{\psi_{2}} + (1-a)\bra{2}\pi_{b}^{\lambda}\ket{2} + \sqrt{a(1-a)}\left(\bra{\psi_2}\pi_{b}^{\lambda}\ket{2}+\bra{2}\pi_{b}^{\lambda}\ket{\psi_2}\right)\right\} \nonumber \ .
\end{align}
The support of the POVM onto the qubit space spanned by the test states $\ket{\tilde{\psi}_0}$ and $\ket{\tilde{\psi}_1}$ is constrained by the reproducibility of the observed statistics $p(b|x)$. However, the support onto the sub-space spanned by $\ket{2}$ does not have any constraints applied. This implies that $p_g$ is maximum whenever the measurement described by the POVM $\left\{\pi_{b}^{\lambda}\right\}$ has minimal support on the constrained subspace. Thus, the upper bound on $p_g$ is given whenever $a$ is minimal, which we know is lower bounded indirectly by \eqref{eq:bound_on_a} whenever the overlaps of the prepared states are also bounded.

On the SDP this is reflected by considering the discrimination of the qutrit states $\ket{\tilde{\psi}_x}$. Assume that Eve is even allowed to change the angles $\phi$ and $\theta$, so that the bounds on the overlaps are still satisfied. Eve can pick them to be the ones she wants in order to make the support onto the third dimension as large as she can. Let's see what is the best she can do. We first relate both angles in a single expression by first writing $a=\left(d_{02}^{2}-d_{12}^{2}\right)/\left(\sqrt{1-d_{01}^{2}}\sin\theta\cos\varphi\right)$ and equating with the right-hand side in \eqref{eq:bound_on_a}. One gets
\begin{equation}
    \cos\varphi = \frac{d_{02}^{2}-d_{12}^{2}}{d_{02}^{2}+d_{12}^{2}}\frac{1+d_{01}\cos\theta}{\sqrt{1-d_{01}^{2}}\sin\theta} \ .
\end{equation}
If one plots $\cos\theta$ vs. $\cos\varphi$, one will see that: if $d_{02}>d_{12}$, $\cos\theta$ is maximal if $\cos\varphi = 1$; if $d_{02}<d_{12}$, $\cos\phi$ is maximal if $\cos\varphi = -1$; and if $d_{02}=d_{12}$, $\cos\theta$ is maximal if $\cos\varphi = 0$. The maximal value of $\cos\theta$ is the same in the three cases, being
\begin{equation}
    \left(\cos\theta\right)_{\text{max}} = \frac{2d_{02}d_{12}-d_{01}\left(d_{02}^{2}+d_{12}^{2}\right)}{d_{02}^{2}-2d_{01}d_{02}d_{12}+d_{12}^{2}} \ ,
\end{equation}
which means
\begin{align}
   \label{eq:tetrahedron} a\geq\frac{d_{02}^{2}+d_{12}^{2}-2d_{01}d_{02}d_{12}}{1-d^{2}_{01}} \ .
\end{align}
Equation \eqref{eq:tetrahedron} defines the surface of a tetrahedron with curved faces. The amplitude $a$ decreases towards the center of the tetrahedron non-symmetrically, as we show in \figref{fig:tetrahedron}.

\begin{figure}
    \centering
    \includegraphics[trim={4cm 1cm 4cm 0},clip,width=\textwidth]{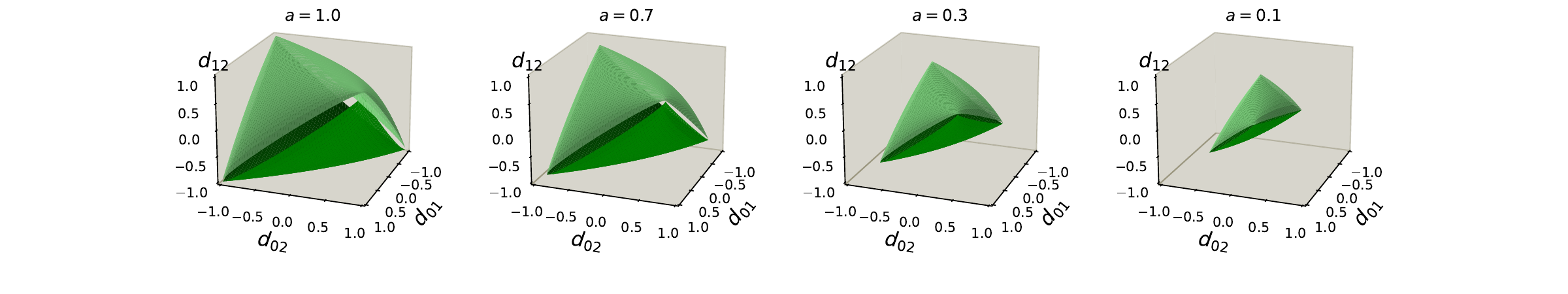}
    \caption{Tetrahedron formed by the available overlap configurations in the preparation of three non-orthogonal quantum states. The parameter $a$ indicates the minimal support of the prepared states onto a two-dimensional subspace if only their overlaps are bounded.}
    \label{fig:tetrahedron}
\end{figure}

On the SDP, the prepared states shall only be expressed in terms of the bounds of the overlaps $d_{xy}$, which can be done by replacing the angles $\phi$, $\theta$ and $\varphi$ with $d_{01}$, $d_{02}$ and $d_{12}$ accordingly. This yields
\begin{align}
    &\ket{\tilde{\psi}_{0}} = \sqrt{\frac{1+d_{01}}{2}}\ket{0} + \sqrt{\frac{1-d_{01}}{2}} \ket{1} + 0 \ket{2} \nonumber \\
    &\ket{\tilde{\psi}_{1}} = \sqrt{\frac{1+d_{01}}{2}}\ket{0} - \sqrt{\frac{1-d_{01}}{2}} \ket{1} + 0\ket{2} \label{eq:tilde_states} \\
    &\ket{\tilde{\psi}_{2}} = \frac{1}{\sqrt{2}}\frac{d_{02}+d_{12}}{\sqrt{1+d_{01}}}\ket{0}+\frac{1}{\sqrt{2}}\frac{d_{02}-d_{12}}{\sqrt{1-d_{01}}}\ket{1} + \sqrt{1-\frac{|d_{02}+d_{12}|^{2}}{2(1+d_{01})}-\frac{|d_{02}-d_{12}|^{2}}{2(1-d_{01})}}\ket{2} \nonumber \ .
\end{align}
A particular choice of bounds on the overlaps and their phases limits the accessibility to a the third dimension by Eve. Concretely, one can tune the states to be symmetric in the sense that $d_{02}=d_{12}^{*}=\tilde{d}e^{i\gamma}$. Thus, fulfilling the relation $\tilde{d}^{2}=(1-d^{2}_{01})/(2(1-d_{01}\cos{2\gamma}))$ makes the third component of $\ket{\tilde{\psi}_{2}}$ null. This means that we can be sure that any potential eavesdropper will gain no information of the outcome by reaching into an additional third dimension.

\section{Specific details on the implementation}
\label{app.implementation}

In this section of the appendix we detail the specific parameters to be adjusted to obtain the desired statistics in the measurement outcomes on the proposed implementation.

The proposed implementation for the USD setup consists in preparing the equi-probable two-mode coherent states $\ket{\psi_{0}}=\ket{\alpha}\otimes\ket{0}$ and $\ket{\psi_{1}}=\ket{0}\otimes\ket{\alpha}$. These can be unambiguously discriminated by means of using only photo detectors in each mode. If only the photo detector in the first mode clicks, that would mean that state $\ket{\alpha 0}$ had been prepared, and thus, we associate the outcome $b=0$. Otherwise, if only the second photo-detector clicks, means that $\ket{0\alpha}$ was prepared and we associate $b=1$ as a measurement outcome. Note that if these two states are prepared, there is a possibility that none of the detectors click. If that happens, the measurement is uncertain of which state was prepared and we associate this events with the inconclusive measurement outcome $b=2$. Assume now that we include the preparation of a third two-mode coherent state $\ket{\psi_{2}}=\ket{\beta_{0}\beta_{1}}$ into play. Whenever this state is prepared, either only one detector can click, the other, none or even both at the same time. Let us go through all possible measurement events and their probabilities to happen whenever a generic state $\ket{\psi_{x}}$ is prepared.
\begin{align}
    1=&\bra{\psi_{x}}\left(\mathds{1}\otimes\mathds{1}\right)\ket{\psi_{x}}=\bra{\psi_{x}}\left(\sum_{n=0}^{\infty}\ket{n}\bra{n}\right)\otimes\left(\sum_{m=0}^{\infty}\ket{m}\bra{m}\right)\ket{\psi_{x}} \nonumber \\
    =&\bra{\psi_{x}}\left(\ket{0}\bra{0}+\sum_{n=1}^{\infty}\ket{n}\bra{n}\right)\otimes\left(\ket{0}\bra{0}+\sum_{m=1}^{\infty}\ket{m}\bra{m}\right)\ket{\psi_{x}} \\
    =&\bra{\psi_{x}}\left(\underset{\text{Detector does't click}}{\underbrace{\ket{0}\bra{0}\otimes\ket{0}\bra{0}}}+\underset{\text{Click on late bin}}{\underbrace{\ket{0}\bra{0}\otimes\sum_{m=1}^{\infty}\ket{m}\bra{m}}}+\underset{\text{Click on early bin}}{\underbrace{\sum_{n=1}^{\infty}\ket{n}\bra{n}\otimes\ket{0}\bra{0}}}+\underset{\text{Click on both early and late bins}}{\underbrace{\sum_{n=1}^{\infty}\ket{n}\bra{n}\otimes\sum_{m=1}^{\infty}\ket{m}\bra{m}}}\right)\ket{\psi_{x}} \nonumber \\
    =&|\braket{00|\psi_{x}}|^{2} + \sum_{m=1}^{\infty}|\braket{0m|\psi_{x}}|^{2} + \sum_{n=1}^{\infty}|\braket{n0|\psi_{x}}|^{2} + \sum_{n=1}^{\infty}\sum_{m=1}^{\infty}|\braket{nm|\psi_{x}}|^{2} \nonumber \ .
\end{align}
The event consisting in a simultaneous click at both time-bins does not come into play when unambiguously discriminating the two-mode coherent states $\ket{\alpha 0}$ and $\ket{0\alpha}$. In fact, that event would correspond in a unambiguous identification of the third state $\ket{\beta_0\beta_1}$. Since our aim is to only consider measurement strategies able to unambiguously discriminate solely states $\ket{\alpha 0}$ and $\ket{0\alpha}$, whenever that event occurs we will consider that $b=0$ with probability $g_0$, $b=1$ with the same probability $g_1$ and the rest of the times $b_2$ with probability $b=2$. No data will be discarded so that the certified randomness is not affected. The considered events and their corresponding probabilities depending on which state was prepared after the post-processing are summarized in table \ref{tab:usd}.
\begin{table}[H]
    \centering
    \begin{tabular}{|c|c|c|c|} \hline
    \backslashbox{Prepared state}{Meas. Event} & $b=0$ & $b=1$ & $b=2$   \\ \hline
      & & & \\ 
     $\ket{\psi_{0}}=\ket{\alpha}\otimes\ket{0}$ & $1-e^{-|\alpha|^{2}}$ & $0$ & $e^{-|\alpha|^{2}}$ \\
      & & & \\
     $\ket{\psi_{1}}=\ket{0}\otimes\ket{\alpha}$ & $0$ & $1-e^{-|\alpha|^{2}}$ & $e^{-|\alpha|^{2}}$ \\
       & & & \\
     $\ket{\psi_{2}}=\ket{\beta_0}\otimes\ket{\beta_1}$ & \begin{tabular}{c}
        $\displaystyle \left(1-e^{-|\beta_0|^{2}}\right)e^{-|\beta_1|^{2}}$  \\
        $\displaystyle +g_0\left(1-e^{-|\beta_0|^{2}}\right)\left(1-e^{-|\beta_1|^{2}}\right)$
     \end{tabular} & \begin{tabular}{c}
        $\displaystyle e^{-|\beta_0|^{2}}\left(1-e^{-|\beta_1|^{2}}\right)$  \\
        $\displaystyle +g_1\left(1-e^{-|\beta_0|^{2}}\right)\left(1-e^{-|\beta_1|^{2}}\right)$
     \end{tabular} & \begin{tabular}{c}
        $\displaystyle e^{-|\beta_0|^{2}}e^{-|\beta_1|^{2}}$  \\
        $\displaystyle +g_2\left(1-e^{-|\beta_0|^{2}}\right)\left(1-e^{-|\beta_1|^{2}}\right)$
     \end{tabular}
     \\ 
       & & & \\ \hline
    \end{tabular}
    \caption{Summary of the considered measurement events and their corresponding re-normalized probabilities for the USD setup.}
    \label{tab:usd}
\end{table}
For simplicity and without loss of generality, we will consider non-imaginary coherent amplitudes only. The overlaps of the prepared states are characterized by the amplitudes of the coherent states as follows
\begin{align}
    d_{01} = e^{-\alpha^{2}} \qquad d_{02} = e^{-\frac{(\alpha-\beta_0)^{2}}{2}}e^{-\frac{\beta_1^{2}}{2}} \qquad d_{12} = e^{-\frac{\beta_0^{2}}{2}}e^{-\frac{(\alpha-\beta_1)^{2}}{2}} \ .
\end{align}
Over a set of runs, we observed that the best and simplest choice is to pick $g_{0}=g_{1}=1/2$ and so $g_{2}=0$.

\section{Finite size effects and entropy accumulation}
\label{app.finite}

In this section we explain how we treat finite size effects from the data extracted in the experiment. Also, we explain how we can abandon the general assumption of independent and identically distributed rounds (i.i.d) though the entropy accumulation theorem (EAT) as is explained in \cite{arnon2018}.

\subsection{Finite-size effects under the i.i.d. assumption: Asymptotic Equipartition Property}

In the real life implementation of the protocol, the observed statistics are built from finite sets of collected data. Thus, the entropy is computed based on a finite number of samples. In order to incorporate such finite-size effects into our analysis, we make use of the asymptotic equipartition property (AEP) generalised to the quantum theory \cite{tomamichel2009}. This allows us to quantify the amount of certifiable randomness over a fixed number of experiment rounds assuming these are independent and identically distributed (i.i.d.). From the experiment we collect pairs of data-points for each question (or state preparation $x$) and answer (measurement outcome $b$). We label by $n_{b,x}$ the number of pairs with question $x$ and answer $b$. Then, from the set of questions we label $n_x$ the number of questions $x$, and form the set of answers we label $n_b$ the number of answers $b$. The total number of extracted data-points is $N=\sum_{b}n_b = \sum_x n_x = \sum_{b,x}n_{b,x}$. From these, we can obtain the observed frequencies as $\text{freq.}(b|x)=n_{b,x}/\sum_{b}n_{b,x}$. Then, we can compute the single round Shannon entropy using the see-saw optimisation method introduced in \appref{ap:seesaw}. The quantum AEP then states that the smooth-min-entropy ($H_{\text{min}}^{\varepsilon}\left(B^{N}|E^{N}\right)$) per-round can be bounded using the following expression
\begin{align}
\label{eq:aep}
\frac{1}{N}H_{\text{min}}^{\varepsilon}\left(B^{N}|E^{N}\right) \geq H\left(B|E\right) - \frac{\delta(\varepsilon,\eta^{\text{AEP}})}{\sqrt{N}} \ ,
\end{align}
where the von Neumann entropy $H\left(B|E\right)$ in our case is reduced to the Shannon entropy per-round, and
\begin{align}
\delta(\varepsilon,\eta^{\text{AEP}}) := 4\log\eta^{\text{AEP}}\sqrt{\log\frac{2}{\varepsilon^{2}}} \quad \text{with} \quad
\eta^{\text{AEP}} \leq \sqrt{2^{-H_{\text{min}}(B|E)}} + \sqrt{2^{H_{\text{max}}(B|E)}} + 1 \ .
\end{align}
To compute the max-entropy $H_{\text{max}}(B|E)$ we use the results from Ref.~\cite{konig2009}. Specifically, for any separable bi-partite state shared between Bob and Eve $\rho_{BE}=\rho_B \otimes \rho_E$, the max-entropy can be obtained as
\begin{align}
H_{\text{max}}(B|E) = 2\log\Tr\left[\sqrt{\rho_B}\right] \ .
\end{align}
Indeed, our assumptions on the state preparations (i.e. bounded overlaps and pure states) fit within this case. We thus have all essential ingredients to compute a bound on the smooth-min-entropy per-round using \eqref{eq:aep} in the i.i.d. case.

\subsection{Entropy accumulation theorem: dropping the i.i.d. assumption}

The i.i.d. assumption is not very attractive in (semi-)device independent protocols. Indeed, this assumes that the eavesdropper cannot learn from past rounds to have a better guess in future rounds (sort of as if the eavesdropper looses its memory in each round). To get rid of this strong assumption, we refer to the entropy accumulation theorem (EAT) generalised to prepare-and-measure scenarios \cite{metger2022} and its application in \cite{metger2023}. Here we adapt it to our semi-device independent prepare-and-measure scenario. The EAT places a bound on the \textit{smooth-}min-entropy, that is the maximum min-entropy of a distribution $\varepsilon$-close to the target distribution, per round in a prepare-and-measure experiment with $N$ rounds. The EAT implies that the operationally total relevant uncertainty about the total set of outcomes over $N$ rounds $B_{1}^{N}$ corresponds to the sum of the entropies of the individual rounds to first order in $N$ under the i.i.d. assumption, plus a contribution from not assuming the i.i.d. case. This contribution is provided given that one quantifies the uncertainty of each individual round with the von Neumann entropy of a suitable chosen state. Formally, the generalised EAT reads
\begin{align}
\label{eq:GEAT}
    \frac{1}{N}H_{\text{min}}^{\varepsilon}(B^{N}|E^{N}) \geq \underset{c_{N}\in\Omega}{\min} f_{min}\left(\text{freq.}(c_{N})\right) - \frac{\alpha-1}{2-\alpha}\frac{\text{ln}(2)}{2}V^2 - \frac{1}{N}\frac{g(\varepsilon)+\alpha\log\left(1/\text{Pr}\left[\Omega\right]\right)}{\alpha-1} - \left(\frac{\alpha-1}{2-\alpha}\right)^{2}K'(\alpha) \ .
\end{align}
Several new elements appear in \eqref{eq:GEAT}, let us properly introduce them one by one. First, $f_{min}$ is a so-called \textit{min-tradeoff function} and it is defined as an analytical function on the observed frequencies (freq.$(c_N)$) computed on the events whenever randomness is certified ($c_{N}$) such that it lower-bounds the minimum entropy over all possible post-measurement states. Namely,
\begin{align}
\label{eq:trade_off}
	f_{min}\left(q\right)  \leq \underset{\nu\in\Sigma_i(q)}{\min} H(B_i|E_i)_\nu
\end{align}
for $\Sigma_i(q)$ being the set of states that can be generated after the measurement at the $i^{th}$ experiment round given the observed frequencies $q$. The lower bound in \eqref{eq:GEAT} is computed on the minimum $f_{min}$ over the total number of observed events $c_{N}$ belonging to a particular chosen event $\Omega$, e.g. the winning condition in non-local games. In our case $\Omega$ represents all events where we certify randomness, i.e. whenever $x=2$, $\forall b$, occurring with probability Pr$(\Omega)=p_{x=2}=1/2$. Here, we take the objective function of the dual semidefinite program in \eqref{eq:dual_sdp} a convenient choice of $f_{min}$. That is,
\begin{align}
f_{min}(\text{freq.}(b|x)) = c_m - \sum_{i=0}^{m-1} \left\{\sum_{b,x}\tilde{\nu}^{i}_{b,x} \ \text{freq.}(b|x) + \Tr\left[\tilde{R}^{i}\right]\right\} \ ,
\end{align}
for $\tilde{\nu}^{i}_{b,x}$ and $\tilde{R}^{i}$ representing feasible points of the dual semidefinite program in \eqref{eq:dual_sdp} for each iteration $i$. For any set of observable frequencies $\text{freq.}(b|x)$, the function $f_{min}(\text{freq.}(b|x))$ is always a lower-bound on the minimum Shannon entropy. Moreover, from the min-tradeoff function we need to compute the maximum $\text{Max}(f_{min})$, minimum $\text{Min}(f_{min})$ and its variance $\text{Var}(f_{min})$. These are then used to compute the following quantities
\begin{align}
g(\varepsilon) =& -\log(1-\sqrt{1-\varepsilon^2}) \ , \\
V =& \log(2 n_{B}^2 + 1) + \sqrt{2+\text{Var}(f_{min})} \ , \\
K'(\alpha) =& \frac{(2-\alpha)^3}{6(3-2\alpha)^3\text{ln}(2)}2^{\frac{\alpha-1}{2-\alpha}(2\log n_{B}+\text{Max}(f_{min})-\text{Min}(f_{min}))}\text{ln}^{3}\left(2^{2\log n_{B}+\text{Max}(f_{min})-\text{Min}(f_{min})}+e^2\right) \ ,
\end{align}
with $n_{B}$ corresponding to the number of measurement outcomes. Finally, the bound in \eqref{eq:GEAT} is originally derived using some appropiate properties of the Reny entropies $H_{\alpha}$ for $\alpha\in\left(1,2\right)$. We thus set $\alpha=1+\mathcal{O}(1/\sqrt{N})$ which is entirely motivated by Ref~\cite{metger2022}, so that one gets a correction term $\mathcal{O}(1/\sqrt{N})$ in the bound on the certifiable entropy per-round.

\subsection{Applicability of the EAT}

In the following we argue that the required conditions on the prepare-and-measure framework we consider are satisfied in our setup in order to apply the entropy accumulation theorem (EAT).

Each round of the prepare-and-measure protocol is defined through a quantum channel $\mathcal{M}_i:R_{i-1} E_{i-1} \rightarrow B_i R_i E_i$. This channel represents the evolution of Bob's internal memory in register $R_{i-1}$ and Eve's information in register $E_{i-1}$ after each run of the experiment, outputting new information on Bob's measurement result $B_i$, in addition to updated information in Bob $R_i$ and Eve's register $E_i$. The generalised entropy accumulation theorem is applicable to all prepare-and-measure scenarios which satisfy the following non-signaling condition,
\begin{align}
    \label{eq:ns_condition}
    \Tr_{B_i R_i} \circ \mathcal{M}_i = \mathcal{R}_i \circ \Tr_{R_{i-1}} \ ,
\end{align}
for $\mathcal{R}_i:E_{i-1}\rightarrow E_i$. This condition implies that the marginal output on Eve's side do not depend on the input on Bob's side (i.e. the local information transferred from the previous round $R_{i-1}$).  Assuming a secure lab, this is well justified in our experiment.

Let us now concentrate to scenarios with classical side information. The evolution $\mathcal{M}_i$ is taken over a shared quantum state between Bob and Eve $\rho_{BE}$. For classical side information, any information that may be leaked over the prepare-and-measure rounds is captured by Bob and Eve sharing a global state which is not entangled, i.e. at step $i$: $\rho^{i}_{B^{i}R_{i}E_{i}}=\sum_{\lambda}q_{\lambda}\rho_{B^{i}R_{i}}^\lambda \otimes \rho_{E_{i}}^\lambda$ for $B^{i}=B_1\ldots B_n$ and $n$ being here the total number of rounds. Additionally, assuming a secure lab implies that Eve does not have access to the prepare-and-measure setup when the experiment is running. This is well justified by the secure environment in which we run our protocol. Added to the assumption of classical side information, this implies that all maps $\mathcal{M}_i$ must be separable, i.e. $\mathcal{M}_i = \sum_\lambda q_\lambda \mathcal{B}_i^\lambda \otimes \mathcal{E}_i^\lambda$, for $\mathcal{B}$ and $\mathcal{E}$ denoting the quantum channels describing the local evolution of Bob and Eve's systems respectively. Therefore
\begin{align}
    \Tr_{B_i R_i} \circ \mathcal{M}_i \left(\rho^{i-1}_{B^{i-1}R_{i-1}E_{i-1}}\right) = \Tr_{B_i R_i} \circ \sum_\lambda q_\lambda \mathcal{B}^{\lambda}_{i}\left(\rho_{B^{i-1}R_{i-1}}^\lambda\right)\otimes \mathcal{E}_{i}^\lambda\left(\rho_{E_{i-1}}^\lambda\right) \\
    = \sum_\lambda q_\lambda \Tr\left[\mathcal{B}^{\lambda}_{i}\left(\rho_{B^{i-1}R_{i-1}}^\lambda\right)\right] \mathcal{E}_i^\lambda\left(\rho_{E_{i-1}}^\lambda\right) \ , \nonumber
\end{align}
where it is clear that setting $\mathcal{E}=\mathcal{R}$ one recovers the non-signalling condition in \eqref{eq:ns_condition}.


\clearpage

\end{widetext}

\end{document}